\def\year{2018}\relax
\documentclass[letterpaper]{article} 
\usepackage{aaai18}  
\usepackage{times}  
\usepackage{helvet}  
\usepackage{courier}  
\usepackage{url}  
\usepackage{graphicx}  
\usepackage{soul}
\usepackage{color}

\usepackage{subfig}
\begin{document}
%
\title{Peer to Peer Hate: Hate Speech Instigators and Their Targets}
\author{Mai ElSherief, Shirin Nilizadeh,\textsuperscript{$*$} Dana Nguyen, Giovanni Vigna, Elizabeth Belding\\
University of California, Santa Barbara\\
\textsuperscript{$*$}CyLab, CMU Silicon Valley\\
{\{mayelsherif, dananguyen, vigna, ebelding\}}@ucsb.edu, shirin.nilizadeh@sv.cmu.edu 
}

\maketitle
\begin{abstract}
While social media has become an empowering agent to individual voices and freedom of expression, it also facilitates anti-social behaviors including online harassment, cyberbullying, and hate speech. In this paper, we present the first comparative study of hate speech instigators and target users on Twitter. Through a multi-step classification process, we curate a comprehensive hate speech dataset capturing various types of hate.  
We study the distinctive characteristics of hate instigators and targets in terms of their profile self-presentation, activities, and online visibility. We find that 
hate instigators target more popular and high profile Twitter users, and that participating in hate speech can result in greater online visibility. We conduct a personality analysis of hate instigators and targets and show that both groups have eccentric personality facets that differ from the general Twitter population. Our results advance the state of the art of understanding online hate speech engagement. 
\end{abstract}

\vspace*{-0.1in}
\section{Introduction}
Social media has become a ubiquitous, powerful communication tool. However, while it has enabled rich, quick information sharing and conversation, it has also facilitated anti-social behavior including online harassment, trolling, cyberbullying, and hate speech. In a Pew Research Center study\footnote{http://www.pewinternet.org/2014/10/22/online-harassment/}, 60\% of Internet users had witnessed offensive name calling, 25\% had seen someone physically threatened, and 24\% witnessed sustained harassment of an individual.

In this paper, we focus on speech that denigrates a person because of their innate and protected characteristics, which is also known as~\emph{hate speech}. While there is no consensus on the definition of hate speech, prior work has shown that people are primarily bullied for their \textit{perceived or actual} ethnicity, behavior, physical characteristics, sexual orientation, class or gender~\cite{silva2016analyzing}. 
Targeting a community or individual because of their immutable or prominent characteristics slowly eradicates feelings of safety and security~\cite{hamm1994conceptualizing,levin2013hate}. 
Prior studies have focused on online hate speech detection~\cite{schmidt2017survey} and characterization, e.g., effect of banning hate speech~\cite{chandrasekharan2017you}; on-the-ground events that 
are triggered by hate speech~\cite{williams2015cyberhate,wired2016insideGoogle,benesch2014countering}; and semi-organized raids by instigators to cripple hate speech detection technology~\cite{hine2017kek}. Despite this work, little is known about online hate speech actors, including hate speech instigators and targets.


We present the first comparative study of online hate speech instigators and targets. We curate a dataset of 27,330 hate speech Twitter tweets and extract 25,278 instigator and 22,287 target accounts. 
Prior work has presented evidence that social media can be used to obtain valuable data that incorporates facets of the virtual and physical worlds of bullying~\cite{xu2012learning}. We choose Twitter because it provides a platform for open discourse and a cross-section of the general public, with 328 million monthly active users in 2017~\cite{statista17twitter}. Our work seeks to answer the following research questions:

\noindent
\textit{\textbf{RQ1:}} How do hate instigator and target account characteristics and online visibility differ from each other and from generic Twitter account holders? 

\noindent
\textit{\textbf{RQ2:}} Are there key personality differences between hate speech instigators, targets and general Twitter users?

Due to the lack of public hate speech datasets that include labeled roles of instigators and targets, we curate our own dataset for what we coin ``\textit{Peer to peer}'' hate speech. This paper presents the following contributions:
\begin{itemize}

\item We present the first comparison of hate instigators, targets and general Twitter users in terms of profile self-presentation, Twitter visibility, and personality traits.
\item We provide a compressed lexicon of Hatebase (the world's largest hate expression repository) for hate speech researchers, comprised of 51 terms likely to result in hate speech content across eight different hate classes. We outline a method of semi-automated classification that could be used for directed explicit hate speech data curation. We curate a dataset of 27,330 hate speech tweets, which we make publicly available for other researchers.\footnote{The lexicon and the dataset are available here: \url{https://github.com/mayelsherif/hate_speech_icwsm18}}
\item We examine the visibility of Twitter users through multi-variant regression models and controlling for variables that can   impact visibility measures. 

\end{itemize}

\newpage
Our study yields multiple important findings. First, hate targets often have older accounts while instigators often have younger accounts.
Compared to general users, both instigators and targets are more active in terms of becoming friends with others, posting tweets, and populating profile content.
Targets include 60\% and 40\% more verified accounts than instigators and general users, respectively. 
Even when controlling for variables that can impact visibility measures, we find that higher visibility and participation in hate are correlated. 
More visible Twitter users (with more followers, retweets and lists) are more likely to become targets of hate. 
Finally, instigators and targets 
share some personality traits such as suspiciousness, low emotional awareness, and high anger and immoderation, which differ from personality traits of the general Twitter user population. 

\section{Related Work}

\textbf{Anti-social behavior.} In 1997, the use of machine learning was proposed to detect  classes of abusive messages~\cite{spertus1997smokey}. 
Cyberbullying has been studied on numerous social media platforms, \textit{e.g.,} 
Twitter~\cite{silva2016analyzing} and YouTube~\cite{dinakar2012common}. 
Other work has focused on detecting personal insults 
and offensive language~\cite{burnap2014hate}. 

\textbf{Hate speech characterization.} 
The characterization and correlation of hate speech with contributing factors has recently received attention. 
Factors include on-the-ground ``trigger'' events, \textit{e.g.,} terrorist attacks~\cite{williams2015cyberhate}, crime~\cite{wired2016insideGoogle}, and news~\cite{hine2017kek}. 

Most closely related to our work are~\cite{chatzakou2017mean,ChatzakouKBCSV17,cheng2017anyone,silva2016analyzing,waseem2016hateful}. 
Chatzakou~\textit{et al.}~\cite{ChatzakouKBCSV17} study the users of tweets with the \#Gamergate hashtag. 
Similar to our results, they found that these users tend to have more friends and followers, and are generally more engaged than random users. 
Chatzakou~\textit{et al.}~\cite{chatzakou2017mean} study the properties of bullies and aggressors and employ supervised machine learning to classify Twitter users into four classes: bully, aggressive, spam, and normal. 
In contrast to their dataset, our dataset is more diverse and not biased towards specific types of hate speech. 
Moreover, we compare the characteristics of hate instigators and the targets of hate from multiple perspectives and show that, even when controlling for features that capture the activity level of the users, both hate instigator and target users  are more likely to get attention on Twitter, \textit{i.e.,} they obtain more followers, are retweeted and listed more.

Alternatively,~\citeauthor{cheng2017anyone} find that prior negative mood and the context of the discussion can combine to double participants' baseline engagement in trolling behavior. 
While the authors only used sentiment analysis to investigate mood, we incorporate a full analysis of the Big Five personality traits. In addition, we study the personality traits of both instigators and targets and compare results to a random sample of general Twitter users. 
Silva~\textit{et al.}~\cite{silva2016analyzing} identified hate target groups in terms of their class and ethnicity on Twitter and Whisper by searching for sentence structures similar to ``I $<$intensity$>$ hate $<$targeted group$>$.'' 
However, we identify the actual accounts of hate targets on Twitter,~\textit{i.e.,} those that are explicitly mentioned by hate instigators. Therefore, our analysis provides a unique lens to analyze characteristics of target accounts.

\vspace{-0.2cm}
\section{Preliminaries}
Waseem~\textit{et al.}~outline a typology of abuse language that differentiates between language directed towards a specific individual or entity (\emph{Directed}) versus a general group of individuals who share a common characteristic, e.g., ethnicity or sexual orientation (\emph{Generalized})~\cite{waseem2017understanding}. Another dimension is whether the abusive language is \emph{explicit}, e.g., contains racial, sexist or homophobic slurs, or \emph{implicit}, which is harder to determine without adding contextual variables to the content. In this work, we study instances of \emph{directed} hate speech that occur between two Twitter accounts. We define the following entities: 
\begin{itemize}
\item A \textbf{hate tweet} is an explicit directed tweet that contains one or more hate speech terms used against a Twitter account holder. An example from our dataset is: 
``\textit{@usr n*gger f*ck u igger n*gger n*gger n*gger}.''\footnote{We replace select vowels with the star (*) character in obscene language.} 
This tweet is explicit because of the word ``n*gger;'' it is directed because it targets a specific account (@usr).\footnote{We anonymize all user mentions by replacing them with \textit{@usr}.}

\item A \textbf{hate instigator (HI)} is a Twitter account that posts one or more hate tweets. 
\item A \textbf{hate target (HT)} is a Twitter account targeted by a hate tweet and explicitly mentioned in the tweet using the mention sign (@), e.g.,~\emph{usr} in our example. We note that role labels are not mutually exclusive in our dataset; a HI account may be a HT in another hate tweet.
\end{itemize}



\section{Data and Methods}
Despite the existence of a body of work dedicated to 
detecting hate speech~\cite{schmidt2017survey}, accurate hate speech detection is still extremely challenging~\cite{cnn2016twitter}. A key problem is the lack of a commonly accepted benchmark corpus for the task. Each classifier is tested on a corpus of labeled comments ranging from a hundred to several thousand.
Another option for collecting a dataset is filtering comments based on hate terms and annotating them. This is  challenging because (i) annotation is time consuming and the percentage of hate tweets is very small relative to the total; and (ii) there is no consensus on the definition of hate speech~\cite{sellars2016defining}. Some work has distinguished between profanity, insults and hate speech~\cite{davidson2017automated}, while other work has considered any insult based on the intrinsic characteristics of the person (e.g. ethnicity,  sexual orientation, gender) to be hate speech related~\cite{warner2012detecting}. 

This annotation process can become even harder for role labeling, i.e., annotating actors as instigators, targets, bystanders~\cite{xu2012learning,faris2016understanding}. This is particularly challenging for social networking APIs that do not provide the whole thread of the conversation but only a random sample of comments, as in the case of the Twitter Streaming API. 
In this work, we adopt a definition of hate speech inspired by Facebook's community standards~\cite{fb2017contrharmhatespeech} and Twitter's hateful conduct policy~\cite{twitter2017hatefulconductpolicy} as ``\textit{direct and serious attacks on any protected category of people based on their race, ethnicity, national origin, religion, sex, gender, sexual orientation, disability or disease}.'' To mitigate the aforementioned challenges, we collect our own explicit Twitter hate speech dataset. We describe our semi-automated detection approach for directed explicit hate speech in the following subsections.

\subsection{Data Collection}
\label{data}

\begin{figure}[!tb]
   \centering
        {\includegraphics[width = 0.45\textwidth, height = 2.5cm]{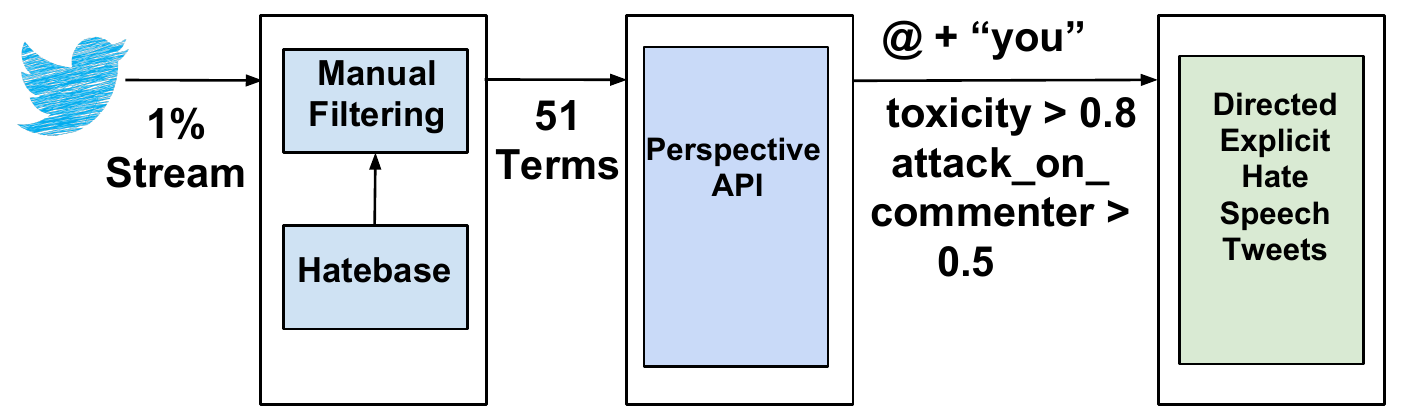}\label{fig:instigators-hist-log}}
   \caption{Flowchart of the filtering process used to obtain our dataset.}
 \label{fig:workflow}
\end{figure}

\textbf{(1) Key phrase-based dataset (HS-1\%):} We adopt a multi-step classification approach. First, we use Twitter's Streaming API\footnote{Twitter Streaming APIs: https://dev.twitter.com/streaming/overview} to procure a 1\% sample of Twitter's public stream from January 1st, 2016 to July 31st, 2017. We began by inspecting hate speech keyphrases in the Hatebase repository\footnote{Hatebase: https://www.hatebase.org/}, the world's largest online repository of structured, multilingual, usage-based hate speech\footnote{We refer to hate speech terms as keyphrases, keywords, hate terms and hate expressions, interchangeably.}. Online users can contribute to Hatebase by adding new derogatory words or phrases, their meaning, and language. Hatebase asks users who add terms to classify the term under one or more of the following hate categories: archaic, class, disability, ethnicity, gender, nationality, religion, and sexual orientation (SexOrient). We use Hatebase as a lexical resource to retrieve English hate terms, broken down as: 42 archaic terms, 57 class, 7 disability, 427 ethnicity, 13 gender, 147 nationality-related, 38 religion, and 9 related to sexual orientation. After careful inspection and five iterations of keyword scrutiny, we removed keyphrases that resulted in tweets with uses distinct from hate speech or phrases that were extremely context sensitive. For example, the word ``pancake'' appears in Hatebase, but clearly can be used in benign contexts. Since our goal was a high quality dataset, we only included key phrases that were highly likely to indicate hate speech. The result is 8, 8, 2, 12, 4, 11, 4, and 2 keyphrases for the above, respective, hate speech classes. Due to the sheer volume of Twitter data, our main focus is to curate a relevant and accurate hate speech dataset with minimal amount of noise.



Despite the qualitative inspection of the keyphrases, when we used the resultant keyphrases to filter tweets from the 1\% public stream, non-hate speech tweets remained in our dataset. To mitigate the effects of obscure contexts and stance 
on the filtering process, we were in need of a hate speech classifier that could remove non-hate speech tweets. Consider the following two tweets: \\
(a): ``\textit{@usr\_1 i'll tear your limbs apart and feed them to the f*cking sharks you n*gger}'' \\
(b): ``\textit{@usr\_2 what influence?? that you can say n*gger and get away with it if you say sorry??}. \\
While both of these tweets contain the word ``n*gger'', the first tweet (a) is pro-hate speech where the hate instigator is attacking \textit{usr\_1}; the second tweet (b) is anti-hate speech in which the tweet author denounces the comments of \textit{usr\_2}. Thus stance detection is vital to consider when classifying hate speech tweets. To mitigate the effects of obscure contexts and stance with respect to hate speech on the filtering process, we used  the Perspective API\footnote{Conversation AI source code: https://conversationai.github.io/} developed by Jigsaw and the Google Counter-Abuse technology team, the model for which is comprehensively discussed in ~\cite{wulczyn2017ex}.\footnote{We also experimented with classifiers including~\cite{davidson2017automated} but found Perspective API to be empirically better.}

The Perspective API contains different models of classification including: toxicity, attack of commenter, inflammatory, and obscene, among others. When a request is sent to the API with specific model parameters, a probability value [0, 1] is returned for each model type. For our datasets, we focus on two models: \texttt{toxicity} and \texttt{attack\_on\_commenter}. The \texttt{toxicity} model is a convolutional neural network trained with word-vector inputs. It measures how likely a comment will make people leave a discussion. The \texttt{attack\_on\_commenter} model measures the probability a comment is an attack on a fellow commenter and is trained on a New York Times dataset tagged by their moderation team. After inspecting the \texttt{toxicity} and \texttt{attack\_on\_commenter} scores for the tweets filtered by the Hatebase phrases, we found that a threshold of 0.8 for \texttt{toxicity} scores and 0.5 for \texttt{attack\_on\_commenter} scores yielded a high quality dataset. 

As a final step to ensure that the resultant tweets attacked a specific Twitter user, we took the remaining tweets in our hate dataset and retained only those tweets that both mention another account (@) and that contain second person pronouns (e.g., ``you'', ``your'', ``u'', ``ur''). The use of second person pronouns has been found to occur with high prevalence in directed hostile messages~\cite{spertus1997smokey}. The result of applying these filters is a high precision hate speech dataset of 27,330 tweets in which HIs use explicit Hatebase expressions against HTs. Figure~\ref{fig:workflow} depicts the filtering process along with our workflow.

\vspace*{0.1in}
\noindent
\textbf{(2) General dataset (Gen-1\%):} 
To provide a larger context for interpretation of our analyses, we compare data from the HS-1\% dataset with a random sample of all general Twitter accounts. To create this dataset, we use the Twitter Streaming API to obtain a 1\% sample of tweets posted per day within the same 18 month collection window and extract the union set of users who posted them. 
We then remove accounts appearing in the HS-1\% dataset, and randomly sample 60K of the remaining users. To mitigate the bias towards more active users, we sample from the union set of users to ensure equiprobable selection of all users, regardless of activity level. While we try our best to remove all the bias, we acknowledge the possibility that this set might include some HIs and HTs. However, later our results show that this bias is likely to have have little impact because we observe significant differences between characteristics of HIs and HTs compared to the general dataset.



\begin{table}[!tb] \centering
 \resizebox{\columnwidth}{!}{%
 \begin{tabular}{@{\extracolsep{5pt}}lccccccc}
\\[-5ex]\hline 
\hline \\[-1.8ex]
& \multicolumn{2}{c}{Total Unique Users} & \multicolumn{2}{c}{Suspended} & \multicolumn{2}{c}{Deleted}\\
\hline \\[-1.8ex] 
HS Type & HI & HT & HI (\%) & HT (\%) & HI (\%) & HT (\%)\\ 
\hline
Archaic & 169 & 169 & 8.3 & 11.2 & 4.1 & 4.1\\
Class & 849 & 837 & 10.0 & 7.3 & 4.9 & 4.4\\
Disability & 8,044& 7,930 & 11.8 & 6.7 & 5.7 & 4.3\\
Ethnicity & 2,073 & 2,045 & 18.8 & 11.3 & 6.6 & 5.2\\
Gender & 13,195 & 13,340 & 9.4 & 5.7 & 5.6 & 4.7\\
Nationality & 78 & 79 & 9.0 & 11.4 & 6.4 & 3.8\\
Religion & 45 & 47 & 13.3 & 19.1 & 13.3 & 2.1\\
SexOrient & 3,638 & 3,584 & 15.3 & 9.0 & 6.9 & 6.0\\
\hline
HS-1\% & 25,278 & 22,857 & 12.8 & 8.3 & 6.5 & 5.7 \\ \hline \\[-1.8ex] 
Gen-1\% & \multicolumn{2}{c}{60,000} & \multicolumn{2}{c}{5.2} &\multicolumn{2}{c}{3.2} \\ \hline
\end{tabular}
}
\caption{Suspended and deleted accounts for all datasets.} \label{users}
\end{table}

Table~\ref{users} shows the number of users in each of our datasets. 
In total, our dataset includes 25,278 hate instigators and 22,857 targets.
The table  shows the quantity of hate tweets for different hate classes. 

The number of keywords used for identifying each class of hate can have an impact on the number of detected HIs and HTs. However, we observe that some classes with fewer keywords, such as \emph{gender}, \emph{disability} and \emph{sexual orientation}, with 4, 2 and 2 keywords, have a higher contribution to our dataset, with 52\%, 32\% and 14\% of HIs. This shows the prevalence of these hate keywords on Twitter. 

Table~\ref{users} also shows the percentages of suspended and deleted accounts. 
The Twitter API returns an error message when the user account is suspended or the user is not found. 
According to Twitter, account suspensions occur when the account is spam, its security is at risk, or it is engaged in abusive tweets or behaviors. 
Twitter accounts that are not found (deleted) occur when the user does not exist. This error could arise for a variety of reasons: the user deactivated their account, the account was permanently deleted after thirty days of deactivation, etc. We label users that no longer exist as \emph{deleted}.
On average, suspended accounts comprise 12.8\% of instigators, 8.3\% of targets, and 5.2\% of general Twitter users.
Additionally, on average, deleted accounts comprise 6.5\% of instigators, 5.7\% of targets, and 3.2\% of general Twitter accounts. Our findings show that instigators and targets are more likely to have their accounts suspended or deleted than general Twitter users, with instigators as the most likely.

Across each hate class, approximately 5\% of accounts are deleted. The only exception is the \emph{Religion} class, where 13\% of hate instigator accounts are deleted. However, this may be the result of the small sample from this class. 
Interestingly, it seems Twitter is more successful in detecting hate  related to \emph{Ethnicity}, \emph{SexOrient} and \emph{Religion} as these categories have the highest number of suspended instigator accounts. 

\begin{figure}[!t]
   \centering
        \subfloat[Tweets by instigators]{\includegraphics[width=0.22\textwidth, height = 2.7cm]{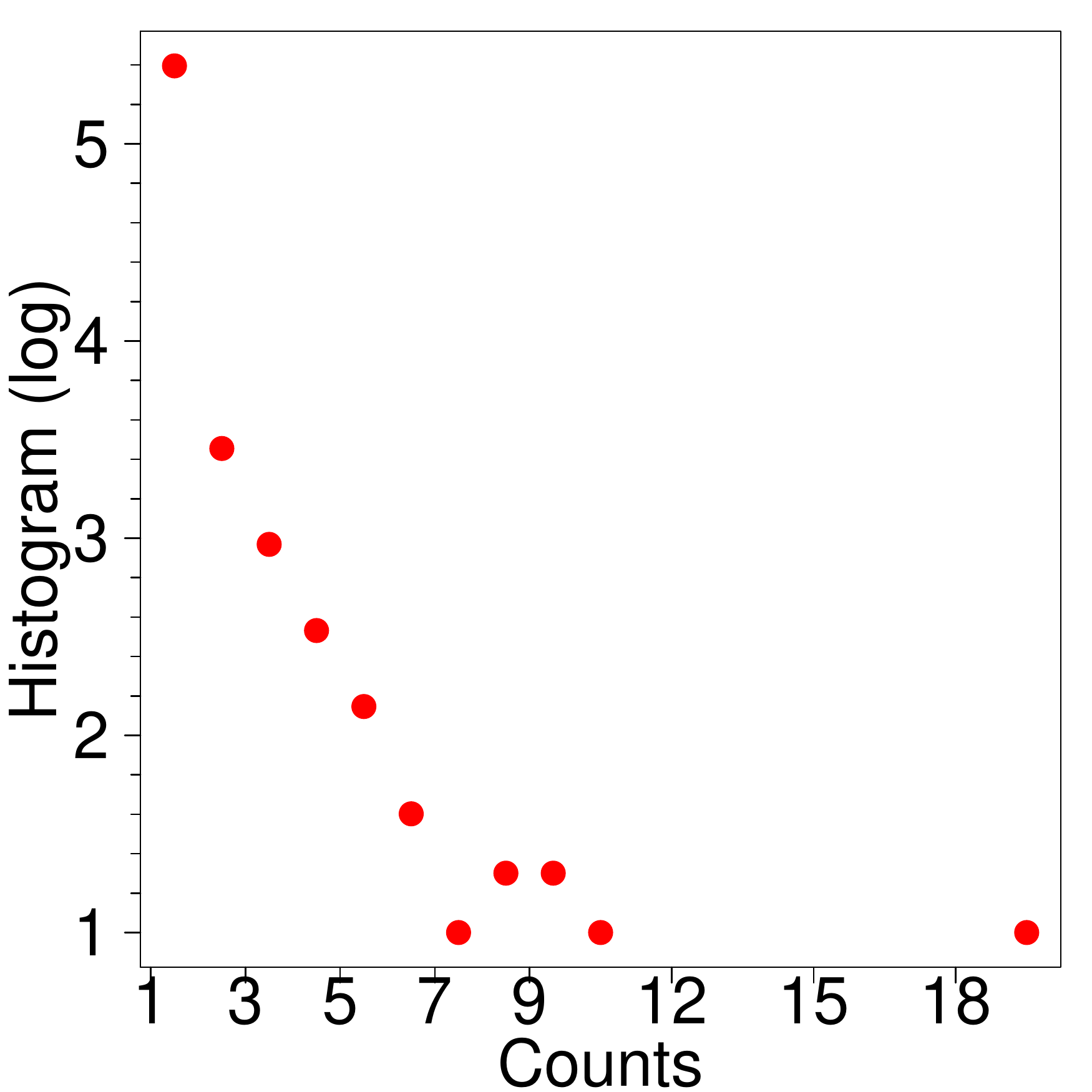}\label{fig:instigators-hist-log}}
       \subfloat[Tweets against targets]{\includegraphics[width=0.22\textwidth, height = 2.7cm]{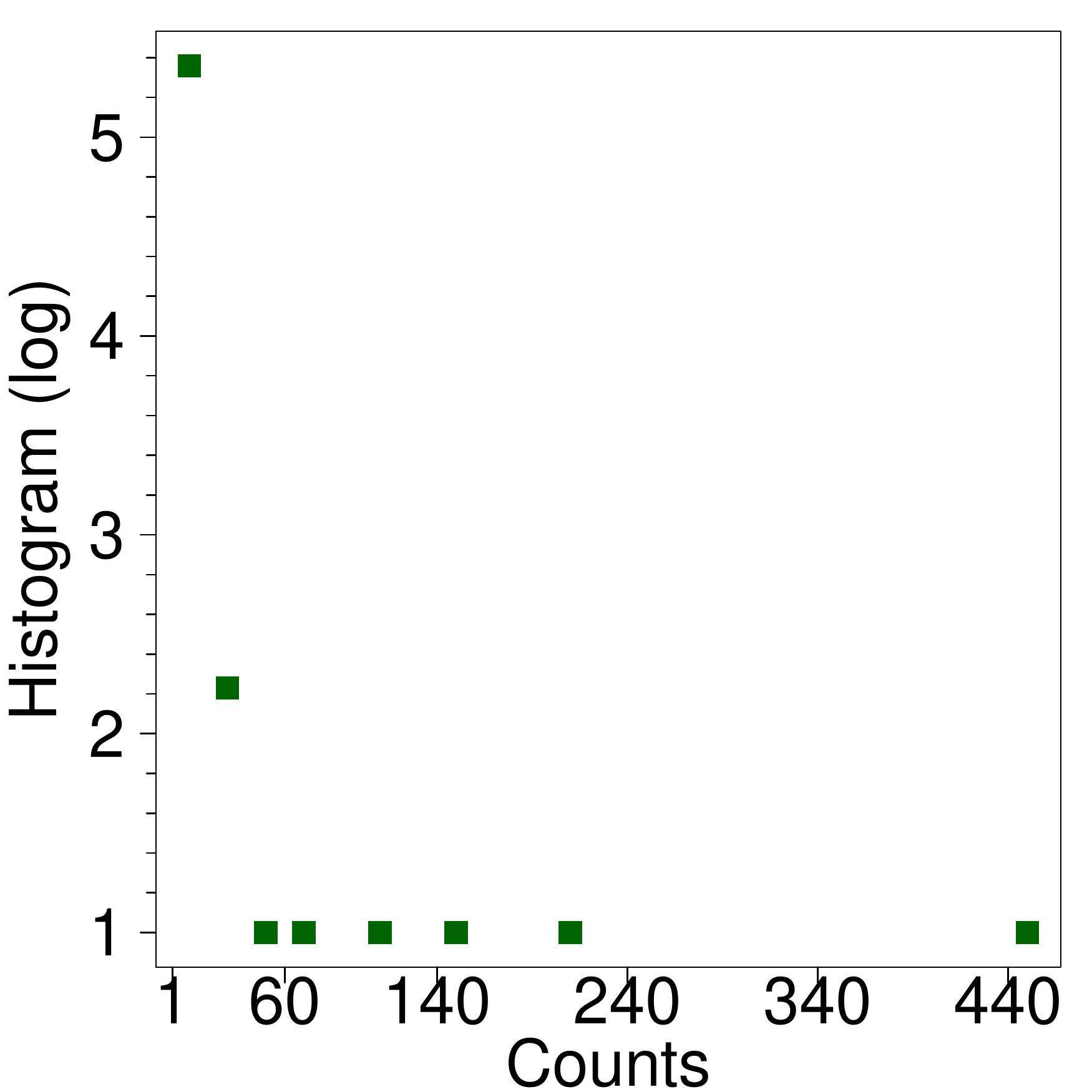}\label{fig:targets-hist-log}}
   \caption{Frequency of hate tweets in HS-1\%. }
 \label{fig:users-hist}
\end{figure}

\begin{table*}[!htbp] \centering 
\tiny
\resizebox{\textwidth}{!}{%
\begin{tabular}{@{\extracolsep{5pt}}lcccc|cccc|cccc} 
\\[-5ex]\hline 
\hline \\[-1.8ex] 
& \multicolumn{4}{c}{Gen-1\% users} & \multicolumn{4}{c}{HIs} & \multicolumn{4}{c}{HTs} \\
\hline \\[-1.8ex] 
Statistic & \multicolumn{1}{c}{Mean} & \multicolumn{1}{c}{Min} & \multicolumn{1}{c}{Max} & \multicolumn{1}{c}{Median}  & \multicolumn{1}{c}{Mean} & \multicolumn{1}{c}{Min} & \multicolumn{1}{c}{Max} & \multicolumn{1}{c}{Median}  & \multicolumn{1}{c}{Mean} & \multicolumn{1}{c}{Min} & \multicolumn{1}{c}{Max} & \multicolumn{1}{c}{Median} \\ 
\hline \\[-1.8ex] 
Followers count& 932 & 0 & 4,589,177 & 93& 1,358 & 0 & 1,006,790 & 259 &  229,676 & 0 & 102,008,153 & 857 \\ 
Friends count&  408 & 0 & 243,937 & 160& 663  & 0 & 1,012,412 & 239& 1,897 & 0 & 1,698,640 & 396\\ 
Tweets count&  4,384 &  0 & 570,550 & 545& 14,160  & 0 & 4,321,652 & 3,266& 29,559 & 1 & 3,644,240 & 10,902 \\ 
Listed count&  8 & 0 & 10,118 & 0& 13 & 0 & 7,855 & 2 & 755 & 0 & 616,271 & 9  \\ 
Retweet counts & 3 & 0 & 13,220 & 0& 30 & 0 & 27,390 & 2 & 623 & 0 & 304,900 & 10\\
Account age (years) & 3.73 & 0.09 & 10.99 & 3.33 & 3.67 & 0.09 & 10.66 & 3.22 & 4.40 & 0.09 & 11.37 & 4.16 \\
len. description (chars) & 45 & 0 & 164 & 28 & 53 & 0 & 164 & 37& 63 & 0 & 164 & 49\\ 
Profile image & 0.95 & 0 & 1 & NA& 0.97 & 0 & 1 & NA& 0.99 & 0 & 1 & NA\\ 
Profile URL & 0.23 & 0 & 1 & NA  & 0.24 & 0 & 1 & NA & 0.40 & 0 & 1 & NA \\  
Geo location & 0.33 &  0 & 1 & NA  & 0.39 &  0 & 1 & NA & 0.51 &  0 & 1 & NA \\  
Location & 0.53 & 0 & 1& NA & 0.61 & 0 & 1 & NA & 0.69 & 0 & 1 & NA \\
Timezone & 0.40 & 0 & 1 & NA & 0.52 & 0 & 1 & NA& 0.68 & 0 & 1 & NA\\ 
Verified & 0.003 & 0 & 1 & NA & 0.002 & 0 & 1 & NA& 0.12 & 0 & 1 & NA\\ 
\hline \\[-1.8ex] 
& \multicolumn{4}{c}{$N = 60,000$} & \multicolumn{4}{c}{$N = 25,278$}&\multicolumn{4}{c}{$N = 22,857$}\\
\hline 
\end{tabular} 
}
  \caption{Descriptive statistics of our datasets.} \label{stats-rnd} 
\end{table*} 

Many account holders in HS-1\% either post more than one hateful tweet, or are hate targets more than once. 
Further, we identify 2,077 (approximately 5\%) accounts that are both hate instigators and targets. 
Figure~\ref{fig:instigators-hist-log} illustrates the logarithmic histogram for the number of hate tweets posted by each instigator account. 
In our HS-1\% dataset, about 10\% of instigator accounts have posted more than one hate tweet. 
Figure~\ref{fig:targets-hist-log} illustrates the histogram representing the number of hate tweets against other accounts. 
Approximately 11\% of accounts are mentioned in more than two tweets, while two specific accounts are mentioned in 449 and 210 hate tweets. 

 \subsubsection{Human-centered dataset evaluation.}
We evaluate the quality of our final dataset by incorporating human judgment using Crowdflower. We provided annotators with a class balanced random sample of $1000$ tweets.\footnote{We used a random sample of $1000$ tweets to keep the monetary cost manageable.} 
To aid annotation, all annotators were provided a set of precise instructions. This included the definition of hate speech according to the social media community (Facebook and Twitter) and examples of hate tweets selected from each of our eight hate speech categories. Then, for each tweet, we asked annotators two questions: (1) whether the tweet is hate speech, and (2) whether the tweet is a direct attack towards the account mentioned in the tweet.
To improve the quality of responses, before assigning a task to annotators, we asked them five test questions with already known responses. If they could not answer at least 80\% of these questions correctly, we identified them as unreliable annotators and removed them from the task.
Each tweet was labeled by at least three independent Crowdflower annotators. 
 

Using the majority vote, we found that annotators labeled 97.8\% of the tweets as hate speech and 94.3\% of tweets as an attack towards the mentioned account. 
We then evaluated the inter-annotator reliability by measuring the agreement percentage of annotators for each of the questions. We found that the agreement percentage for the first question is 92.8\%, and for second question is 82.6\%. These results shows that our hate speech dataset is reliable with minimal noise.
\vspace{-0.2cm}
\subsection{Measures}
We adopt several measures based on prior work to answer our research questions. To compare the account characteristics of HIs and HTs, we investigate whether users have a profile image, set a geo-location and a timezone, whether the account is verified, and the length of the profile description. We study the number of tweets and retweets, friends, followers, and whether the account is enlisted. 
Similar to Nilizadeh~\textit{et al.}~\cite{ICWSM1613003}, we differentiate accounts by \textit{perceived}, as opposed to \textit{actual}, user characteristics.  This is because we can only study how an account holder chooses to represent him/herself, i.e., through a profile photo, and cannot determine their actual characteristics.

We predict user gender by extracting first names and comparing them with those listed in the 1900 -- 2013 U.S. Census~\cite{ICWSM112816,ICWSM1613003}. 
We leverage the IBM Watson Personality Insights API~\cite{ibm2017personalityinsights} to quantify the Big Five personality traits for HIs and HTs. The API has been used in prior studies to correlate personality traits with information-spreading~\cite{lee2014will} and targeted advertising~\cite{chen2015making}.

\section{Analysis}

\subsection{RQ1: Account Characteristics}
\label{rq1}
Our first objective is to understand the differences of self presentation through profile configurations, activity level, and interaction with other users. 
To study profile presentation, we analyze whether profile image, location, and timezone are provided by the user; whether the user has enabled the geo-location to be posted along with their tweets; whether the account is verified by Twitter; and the length of profile description in characters.   
For user activity level, we analyze number of tweets, friends, followers, lists, and retweets. 
The last three of these indicate how Twitter users interact with an account and are used as visibility measures~\cite{ICWSM1613003,Sharma:2012}. 

All characteristics can be extracted from the meta-data provided with the tweets, except the retweet count. 
For every user, we count the number of times the user's tweets are reposted in our 1\% dataset. 
Although the obtained retweet counts only represent a subset of the actual retweets, they provide useful insight when comparing different samples. 

We determine the gender of users by extracting first names and comparing them with first names listed in the U.S. Census dataset obtained from 1900 -- 2013~\cite{ICWSM112816}. 
Some first names are gender-neutral, such as ``Pat.'' 
Similar to other work~\cite{ICWSM112816}, if a name has a female-to-male ratio larger than 0.95 or smaller than 0.05, we label it as female or male; other names
are labeled as `gender ambiguous'.
We are able to extract first names for 53\% of HIs, 55\% of HTs and 56\% of general users. 
HIs use pseudonyms more than others, which can be an indication of desire to hide their identities. 
25\%, 23\% and 8\% of users in the Gen-1\% dataset; 35\%, 10\% and 8\% of users in the instigator dataset;  and 35\%, 12\% and 8\% of users in the instigator dataset are male, female and gender ambiguous, respectively. Instigator and target datasets include 10\% more male and 13\% fewer female users than the Gen-1\% dataset, which implies that \emph{users with female account names are less engaged in hate discussions.}  

Table~\ref{stats-rnd} statistically describes the users in our Gen-1\% and HS-1\% datasets. 
Since the distribution of most characteristics is skewed, in addition to mean, the table also shows the min, max and median of values. 
The table illustrates multiple differences between user types.
The t-tests for account age (by year) suggest that, on average, the accounts for HTs are older than those of HIs ($\mu=4.40$, vs. $\mu=3.67$) ($t=32.18$, $p<0.001$) and generic random users ($\mu=4.40$, vs. $\mu=3.73$) ($t=32.91$, $p<0.001$). Also, the accounts for HIs are younger than those of general random users ($\mu=3.67$ vs. $\mu=3.73$) ($t=3.33$, $p<0.001$).
We observe that compared to random users, HIs and HTs are more active in becoming friends with others, posting tweets, and providing more content on their profiles.

The t-tests for profile description length (in characters) show that, on average, the descriptions provided by HTs are longer than those for HIs ($\mu=63$, vs. $\mu=53$) ($t=20.14$, $p<0.001$). 
The descriptions provided by hate targets and instigators are longer than those of generic random users ($\mu=63$, vs. $\mu=45$) ($t=40.04$, $p<0.001$), ($\mu=53$, vs. $\mu=45$) ($t=19.56$, $p<0.001$). 
These results may suggest that both HIs and HTs are more willing to present themselves.

Table~\ref{chi-square} shows the results of Chi-square tests for the binary variables. In general, HTs reveal more information on their profiles; they are more likely to add image, URL, location and timezone to their profiles compared to both HIs and general Twitter users. There is only one exception where the difference between the distribution of geo-location for HIs and that of HTs is not significant ($p=0.06$). 

Twitter verifies accounts that are of public interest. When accounts are verified, a blue badge appears next to the user's name on their profile.\footnote{Request to verify an account: \\https://support.twitter.com/articles/119135\#}
Interestingly, when comparing HIs and HTs, we observe that HTs include significantly more high profile and established users; 12\% belong to verified accounts. However, HIs themselves are less likely to have verified accounts, even compared to random general users. 

\begin{table}[!tb] \centering 
\scriptsize
 \resizebox{\columnwidth}{!}{%
\begin{tabular}{@{\extracolsep{5pt}}lccccccc} 
\\[-5ex]\hline 
\hline \\[-1.8ex] 
& & \multicolumn{2}{c}{ HT vs. HI} & \multicolumn{2}{c}{Gen-1\% vs. HT} & \multicolumn{2}{c}{Gen-1\% vs. HI} \\
\hline \\[-1.8ex]
 &  \multicolumn{1}{c}{df} & \multicolumn{1}{c}{$X^2$} & \multicolumn{1}{c}{p} & \multicolumn{1}{c}{$X^2$} & \multicolumn{1}{c}{p} & \multicolumn{1}{c}{$X^2$} & \multicolumn{1}{c}{p} \\  
\hline \\[-1.8ex] 
Profile image  & 1 & 7633 & *** & 672& *** & 4901  & *** \\ 
Profile URL  & 1 & 325 & *** & 1858  & *** & 3546  &  ***\\ 
Geo location  & 1 & 3.53  & 0.06 & 1937 & *** & 1801  & ***\\  
Location  & 1 &  1606  & ***& 1389 & ***& 66&  ***\\
Timezone  & 1 & 1389 & *** & 4444  & *** & 797 & ***\\ 
Verified  & 1 & 99  & *** & 6226 & *** & 4789 & ***\\ 
Gender (name) & 1 & 1318 & *** & 1230 & *** & 21 & ***\\
Invalid image & 1 & 2,088,900 & *** & 1,221 & *** & 4,827,400 & ***\\ 
Detected face & 1 & 1,138,200 & *** & 505 & *** & 1,821,700 & ***\\ 
Multiple faces & 1 & 282,530 & *** & 127 & *** & 368,000 & ***\\ 
One face (Male) & 1 & 289,160 & *** & 24,493 & *** & 224,900 & ***\\ 
One face (Female) & 1 & 270,580 & *** & 197,900 & *** & 933,780 & ***\\
\hline \\[-1.8ex]
\multicolumn{5}{l}{Note: *$p<0.05$, **$p<0.01$, ***$p<0.001$}\\
\hline
\end{tabular} 
}
  \caption{Pearson's Chi square tests.}   \label{chi-square} 
\end{table}

Next, we examine the activity and visibility levels of account holders. 
We compare these variables by using Mann-Whitney U tests, because they do not follow a normal distribution. These results are provided in Table~\ref{mann-whitney}. 
Interestingly, HTs have more friends and post more tweets than both HIs and general users. 
They also have higher visibility and influence; their median numbers of followers and retweets are larger than those of both HIs and general users.  

Twitter's `List' feature allows users to organize others by creating topical user lists. 
If some users are known for something, \textit{e.g.,} are 
computer scientists, then they might be listed by others in ``Computer Scientists'' list. Organizing Twitter users into lists helps track  tweets from  those in the list. 
Our results show that targets of hate are listed more often. 
\begin{table}[!tb] \centering 
\scriptsize
 \resizebox{\columnwidth}{!}{%
\begin{tabular}{@{\extracolsep{5pt}}lcccc} 
\\[-5ex]\hline 
\hline \\[-1.8ex] 
 & \multicolumn{1}{c}{ U (HT vs. HI)} & \multicolumn{1}{c}{U (Gen-1\% vs. HT)} & \multicolumn{1}{c}{U (Gen-1\% vs. HI)} & \multicolumn{1}{c}{p} \\
\hline \\[-1.8ex]
Followers  & 321,900K& 183,400K & 504,620K & *** \\ 
Tweets  & 294,930K &  190,920K & 445,380K & ***\\ 
Friends  & 278,670K & 316,970K & 586,540K &  *** \\ 
Lists & 305,450K & 221,840K & 503,890K & *** \\
Retweets & 304,560K & 139,270K & 369,650K &***\\
\hline \\[-1.8ex] 
\end{tabular} 
}
  \caption{Mann-Whitney U tests.}   \label{mann-whitney} 
\end{table}

\begin{figure*}[h]
    \centering
        \subfloat[Followers count]{\includegraphics[width=0.2\textwidth]{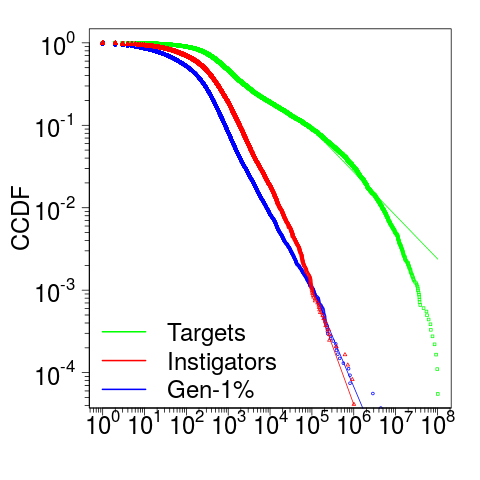}\label{fig:followers-count}}
        \subfloat[Friends count]{\includegraphics[width=0.2\textwidth]{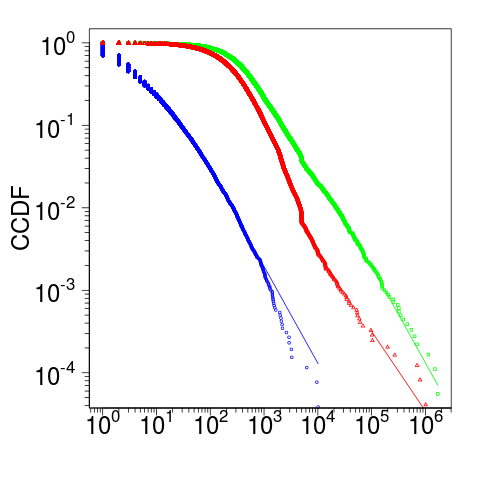}\label{fig:friends-count}}
         \subfloat[Listed count]{\includegraphics[width=0.2\textwidth]{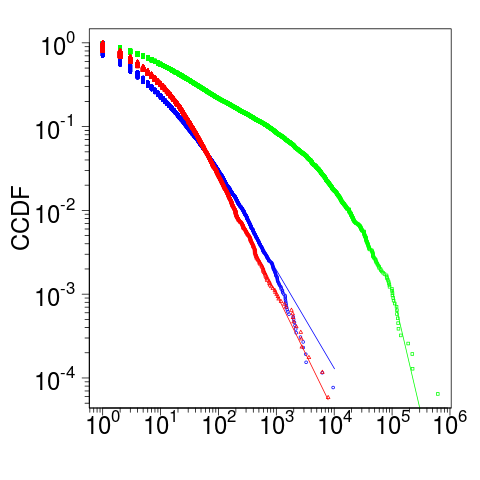}\label{fig:listed-count}}
         \subfloat[Retweets count]{\includegraphics[width=0.2\textwidth]{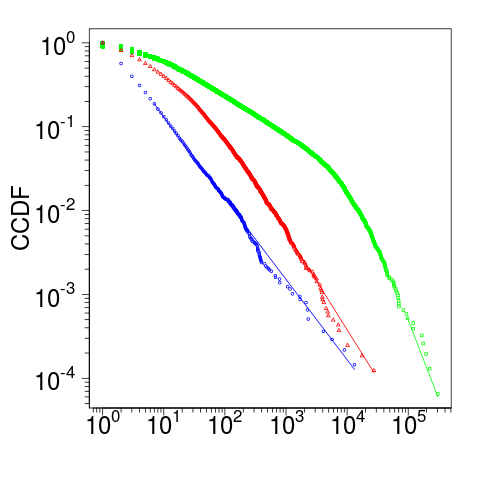}\label{fig:retweeted-count}}
          \subfloat[Tweets count]{\includegraphics[width=0.2\textwidth]{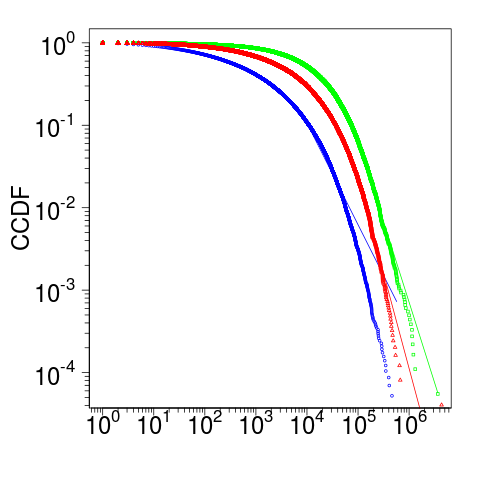}\label{fig:ntweets-count}}
          \caption{Comparison of account characteristics of HIs, HTs, and general users.}
  \label{fig:ccdf}
\end{figure*}

Figure~\ref{fig:ccdf} compares the distribution of the activity and visibility  characteristics of HIs and HTs with those from the Gen-1\% dataset. 
This figure shows CCDF plots for variables that exhibit heavy-tailed distributions.  
Figure~\ref{fig:followers-count} shows that HTs on average have more followers than both HIs and general Twitter users, while the distribution of followers count for HIs is more similar to that of general Twitter users. 
Specifically, the difference between HTs and others is more significant for visibility measures including followers, lists and retweet counts.

\textbf{Visibility:} 
We next examine the visibility of HIs and HTs by controlling for variables that can have an impact on the visibility measures. 
For example, older accounts have had more time to accumulate followers; following many others usually yields more followers by sheer reciprocity; and posting many tweets can increase the chances to be noticed. 
Thus, we incorporate the following control variables in our models: account age, number of tweets, number of friends, and profile characteristics such as URL, location, image, length of user description, timezone and verified, as well as perceived user gender. 
We control for profile characteristics and gender because user self-presentation can affect the way people perceive them, and therefore, can have an impact on visibility measures~\cite{ridgeway2001gender,ICWSM1613003}. 

We select three dependent variables as the main measures of online visibility on Twitter: `number of followers', `retweets,' and `lists.' 
We apply multiple multivariate regression models and present the results from our Poisson regression model. Linear and negative binomial regression models show qualitatively consistent results, although a couple did not converge.

Since our dependent variables exhibit a skewed distribution, examining the whole  population may not capture more nuanced patterns~\cite{yu2003quantile}. 
For example, in Table~\ref{stats-rnd}, we observe that a hate target account holder has more than 100M followers and this user alone can impact the overall and average statistical results. Thus, we adopt the quartile regression technique to analyze our dataset in each quartile. 
We divide the data into quartiles based on each dependent variable and apply multivariate regression models.
Although we include control variables in all models, for brevity, we omit them from the result tables; full tables are available upon request. 
We add followers count as a control for the retweets and lists count models because  more followers may result in being retweeted and listed more. We add lists count as a control for the retweets count model because being listed by many people may result in being retweeted more. 
We report Incident Rate Ratios (IRRs), the exponentiated coefficients of Poisson regressions, which allow us to compare the rates of variables between HIs, HTs, and general users.
 
\begin{table}[!tb] \centering 
\tiny
\resizebox{\columnwidth}{!}{%
\begin{tabular}{@{\extracolsep{5pt}}lccccc} 
\hline 
\hline \\[-1.8ex] 
 & \multicolumn{5}{c}{\textit{Followers count}} \\ 
\cline{2-6} 
\\[-1.8ex] & Poisson & 0.25 Qrt. & 0.5 Qrt. & 0.75 Qrt. & 1.00 Qrt. \\ 
\cline{2-6} \\[-1.8ex]
HT & 2.68$^{***}$ & 0.41$^{***}$ & 0.10$^{***}$ & 0.05$^{***}$ & 2.36$^{***}$ \\ 
IRRs & 14.64 & 1.51 & 1.11 & 1.05 & 10.60\\
\hline \\[-1.8ex] 
 & \multicolumn{5}{c}{\textit{Lists count}} \\ 
\cline{2-6} \\[-1.8ex] 
 HT & 1.93$^{***}$ & 0.04 & 0.08$^{***}$ & 0.06$^{***}$ & 1.59$^{***}$ \\ 
IRRs & 6.92  & 1.036  & 1.08 & 1.06 & 4.92\\
\hline \\[-1.8ex] 
 & \multicolumn{5}{c}{\textit{Retweet count}} \\ 
\cline{2-6}  \\[-1.8ex] 
 HT & 4.06$^{***}$ & 2.57$^{***}$ & 4.18$^{***}$ & 3.76$^{***}$ & 3.35$^{***}$ \\ 
IRRs  & 57.94 & 13.00 & 65.01 & 42.98 & 28.53 \\ 
\hline \\[-1.8ex] 
\end{tabular}
}
  \caption{HTs vs. All Poisson Regressions.}   \label{visibility:targets-all} 
\end{table} 
\begin{table}[!t] \centering 
\tiny
\begin{tabular}{@{\extracolsep{5pt}}llll} 
\hline 
\hline \\[-1.8ex] 
\\[-1.8ex] & \textit{Followers count} & \textit{Lists count} & \textit{Retweet count} \\ 
\cline{2-4} \\[-1.8ex]
 HT (IRRs) & 2.03$^{***}$ (7.65) & 1.59$^{***}$ (4.90) & 3.15$^{***}$ (23.32)\\ 
\hline \\[-1.8ex] 
\end{tabular}
\caption{HTs vs. HIs Poisson Regressions.}   \label{visibility:targets-instigators} 
\end{table} 
\begin{table}[!t] \centering 
\tiny
\begin{tabular}{@{\extracolsep{5pt}}llll} 
\hline 
\hline \\[-1.8ex] 
\\[-1.8ex] & \textit{Followers count} & \textit{Lists count} & \textit{Retweet count} \\ 
\cline{2-4} \\[-1.8ex]
 HI (IRRs) & 0.46$^{***}$ (1.59)& 0.49$^{***}$ (1.62) &1.98$^{***}$ (7.26)\\ 
\hline \\[-1.8ex] 
\end{tabular}
  \caption{HIs vs. Gen-1\% Poisson Regressions.}   \label{visibility:Instigators-general} 
\end{table} 

Table~\ref{visibility:targets-all} shows the results of Poisson regression 
comparing HTs vs. the union of HIs and general users. 
The first column shows the result for the entire sample such that HTs have significantly more followers, are listed and retweeted more than all other users ($p <0.001$). 
Particularly, for followers, lists and retweet counts, the HTs have IRRs 14.64, 6.92 and 57.94 times of those of the union of HIs and general users. Table~\ref{visibility:targets-instigators} illustrates that these findings hold even when HTs are compared only with HIs ($p <0.001$). 
\emph{These results suggest that regardless of user activity level, profile self-presentation, and gender, more visible Twitter users (with more followers, lists, and retweets) are more likely to become target of hate.}

Table~\ref{visibility:Instigators-general} demonstrates the results of models for HIs vs. general users. 
The coefficients for both overall and quartiles models are positive and larger than one, which indicate that HIs are positively associated with being visible. 

In Table~\ref{visibility:targets-all}, quartile regression reveals that the overall and average results are not just the effects of most visible users, and in each quartile, the HTs are more visible than HIs and general users. 
Although the effect of HTs (IRR) increases as one moves from the least visible to most visible users, in almost all quartiles values are larger than one. 
For brevity, we do not report the results of models per quartiles in Tables~\ref{visibility:targets-instigators} and~\ref{visibility:Instigators-general} although the interpretation of their results is consistent with those reported for Table~\ref{visibility:targets-all}.

Comparing the IRR results with those in Tables~\ref{visibility:targets-all} and~\ref{visibility:targets-instigators} shows that the differences between the HTs and HIs are significantly higher than those of HIs and general users. 
\emph{These results also suggest that participating in hate speech and being more visible and popular are related; even when controlling for all mentioned independent variables, both HIs and HTs are more popular and visible than general users.} 

\subsection{RQ2: Personality Traits}
To study the key differences between the personalities of HIs, HTs, and the general population, we use the Twitter REST API to fetch tweet traces of users. A Twitter user can share content on their profile in three different ways: an original tweet, a reply to a tweet written by another user, or a redistribution of a tweet written by another account (retweeting). Retweets do not necessarily indicate content endorsement but suggest content to be viewed by the retweeter's network. Since retweeting content might not reflect the author's point of view, we only include original tweets and replies as part of our personality analysis.
We attempt to fetch the most recent 2000 tweets (excluding retweets) for each account.
We use IBM Watson Personality Insights API\footnote{https://www.ibm.com/watson/services/personality-insights/} for our personality analysis. Since the Personality Insights API requires a minimum of 600 words to obtain statistically significant result estimates, we discard any accounts that do not satisfy this requirement. After discarding suspended and deleted accounts, accounts with statistical insignificance, and accounts with languages other than English, we were able to fetch tweets for a total of 17,951 unique HIs, 17,553 unique HTs, and 12,900 unique general users (pulled from Gen-1\%).\footnote{All sampling errors in our results are less than 0.1.} We use the general users personality results as a means of account sample representation on Twitter. The word count distribution is ($\mu = 11,045.6, \sigma = 7,230.5$) for HI accounts, ($\mu = 12,316.1, \sigma = 7,308.7$) for HT accounts, and ($\mu = 8,108.2, \sigma = 7,288.7$) for accounts in Gen-1\%.

\begin{table*}[!ht] \centering 
\tiny
\resizebox{\textwidth}{!}{%
\begin{tabular}{lccccccccc|cccc}
\\[-5ex]\hline 
\hline \\[-1.8ex] 
& \multicolumn{3}{c}{Medians}  & \multicolumn{2}{c}{ HI vs. HT} & \multicolumn{2}{l}{HI vs. Gen-1\%} & \multicolumn{2}{l}{HT vs. Gen-1\%} & \multicolumn{3}{c}{Hellinger distances}\\ \hline
Personality facet & HI & HT & Gen-1\% & U &p& U &p& U &p& HI-HT & HI-Gen-1\% & HT-Gen-1\% \\ \hline

Agreeableness & 0.06 & 0.1 & 0.4 & 134,790K & *** & 47,512K & *** & 61,130K & *** & 0.11 & 0.37 & 0.27 \\ 

Openness  & 0.49 & 0.51 & 0.5 & 152,400K & *** & 114,760K & 0.18 & 115,840K & *** & 0.03 & 0.03 & 0.04 \\ 

Emotional range  &0.18 & 0.22 & 0.38 & 142,360K & *** & 77,917K & *** & 87,490K & *** & 0.08 & 0.22 & 0.15 \\ 

Conscientiousness  & 0.02 & 0.05 & 0.31 & 128,370K & *** & 35,667K & *** & 55,020K & *** & 0.18 & 0.46 & 0.31 \\ 

Extraversion  & 0.23 & 0.31 & 0.47 & 149,410K & *** & 83,693K & *** & 88,067K & *** & 0.04 & 0.17 & 0.13 \\ 

\hline
\multicolumn{13}{l}{Note: *$p <0.05$ **$<0.01$ ***$<0.001$}\\
\hline
\end{tabular} 
}
 \caption{Scores and Hellinger distances for the Big Five personality traits of HIs, HTs and general users. } 
  \label{table:personality-facets}
\end{table*}

\begin{figure*}[!htb]
    \centering
        {\includegraphics[width=0.19\textwidth, height = 3cm]{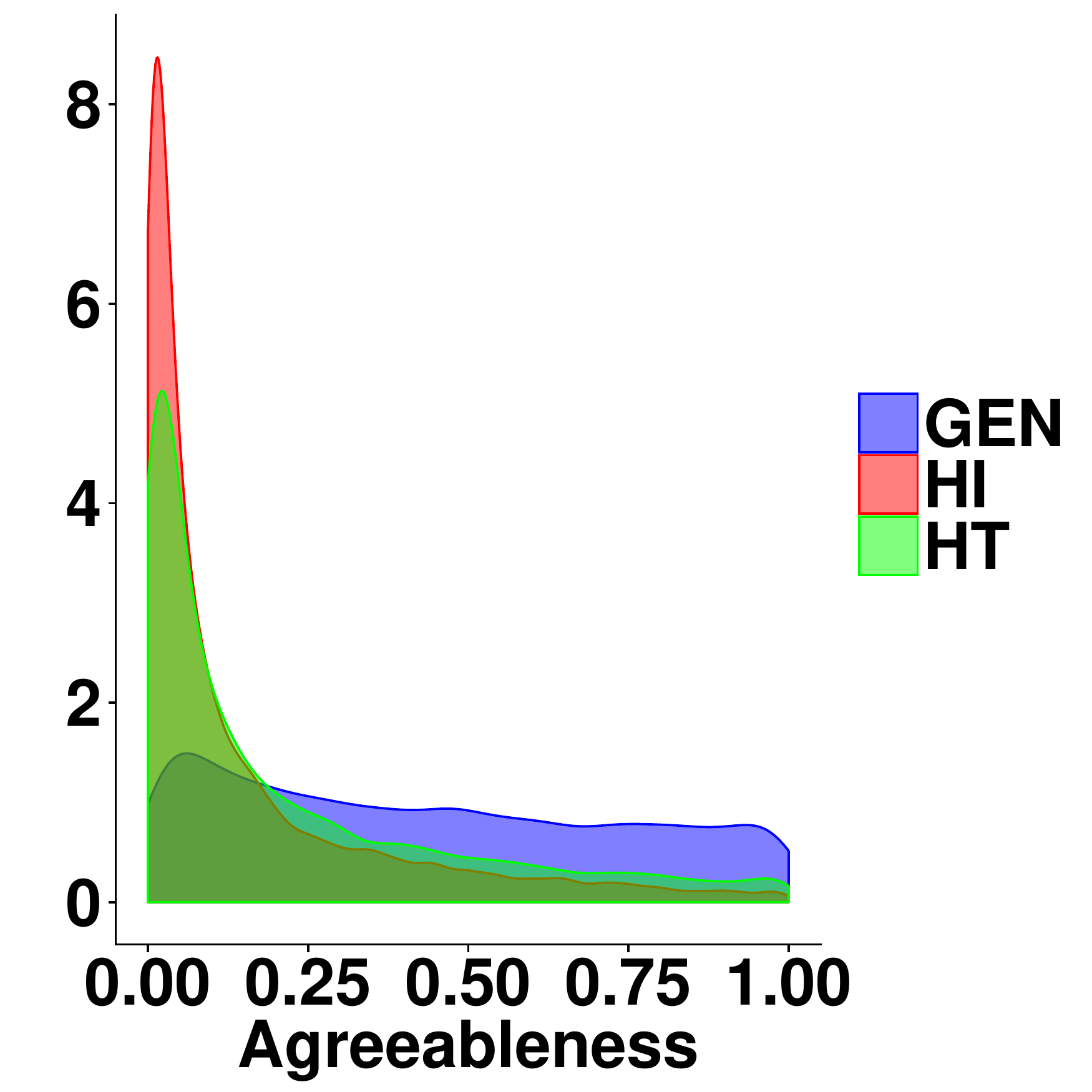}}
         {\includegraphics[width=0.19\textwidth, height = 3cm]{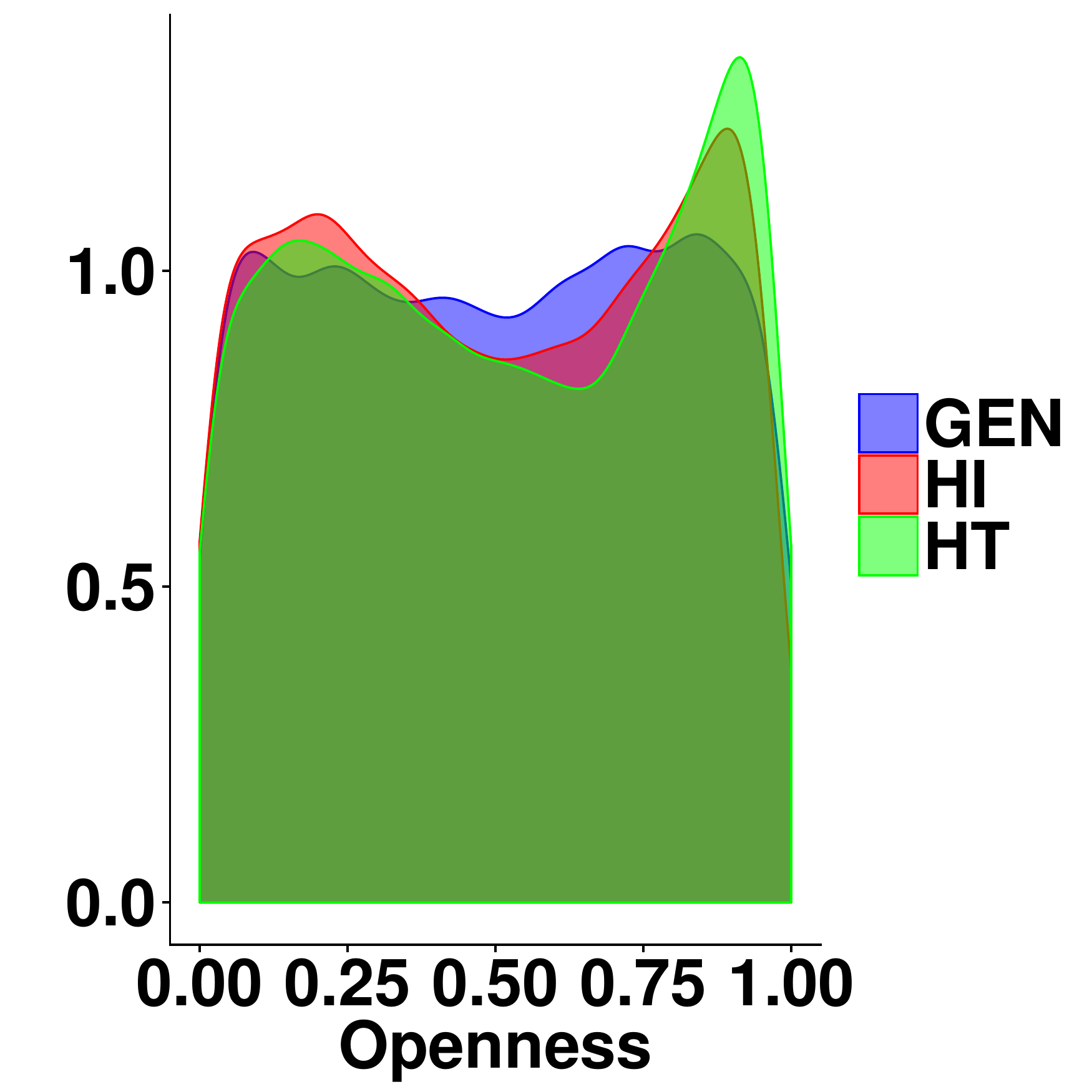}}
         {\includegraphics[width=0.19\textwidth, height = 3cm]{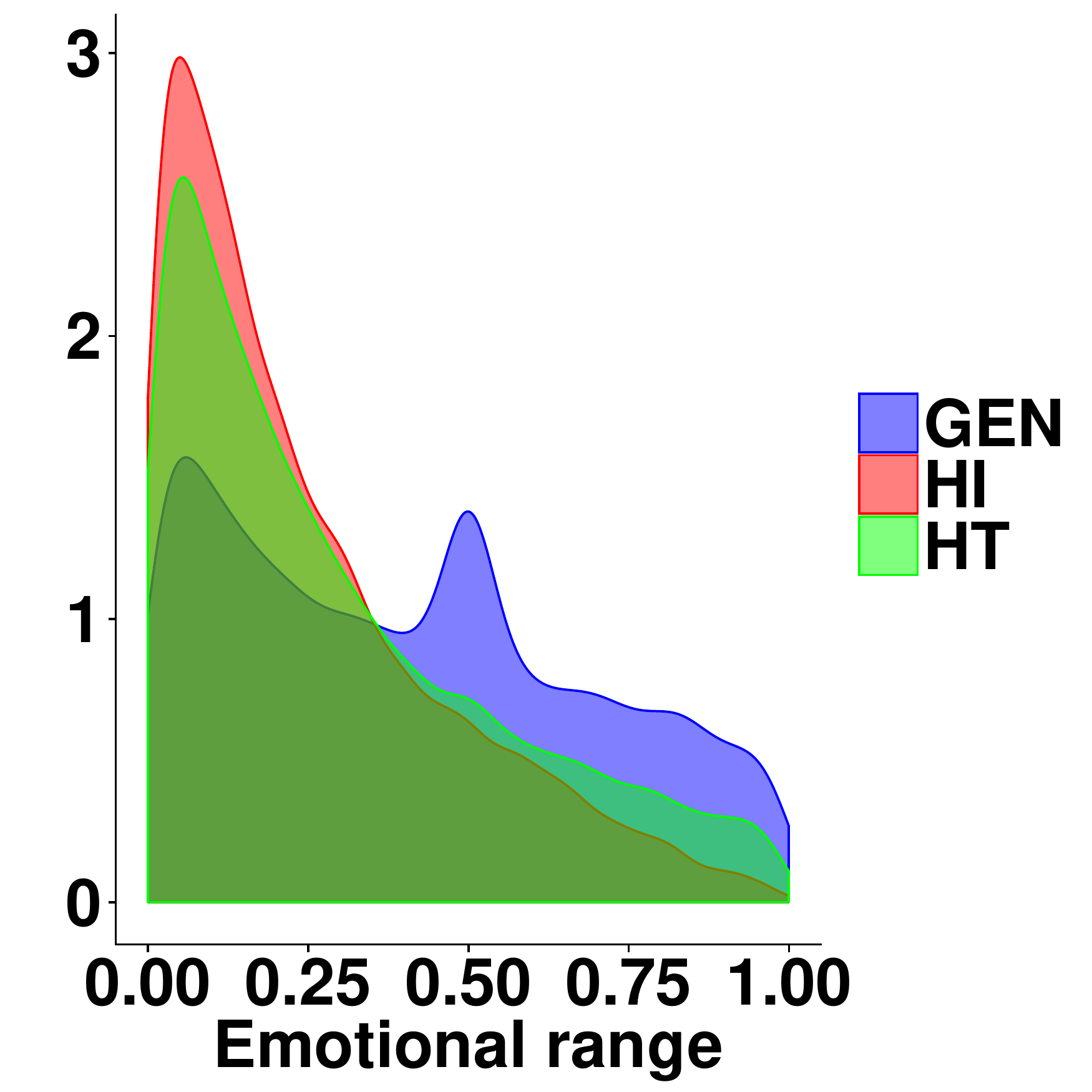}}
        {\includegraphics[width=0.19\textwidth, height = 3cm]{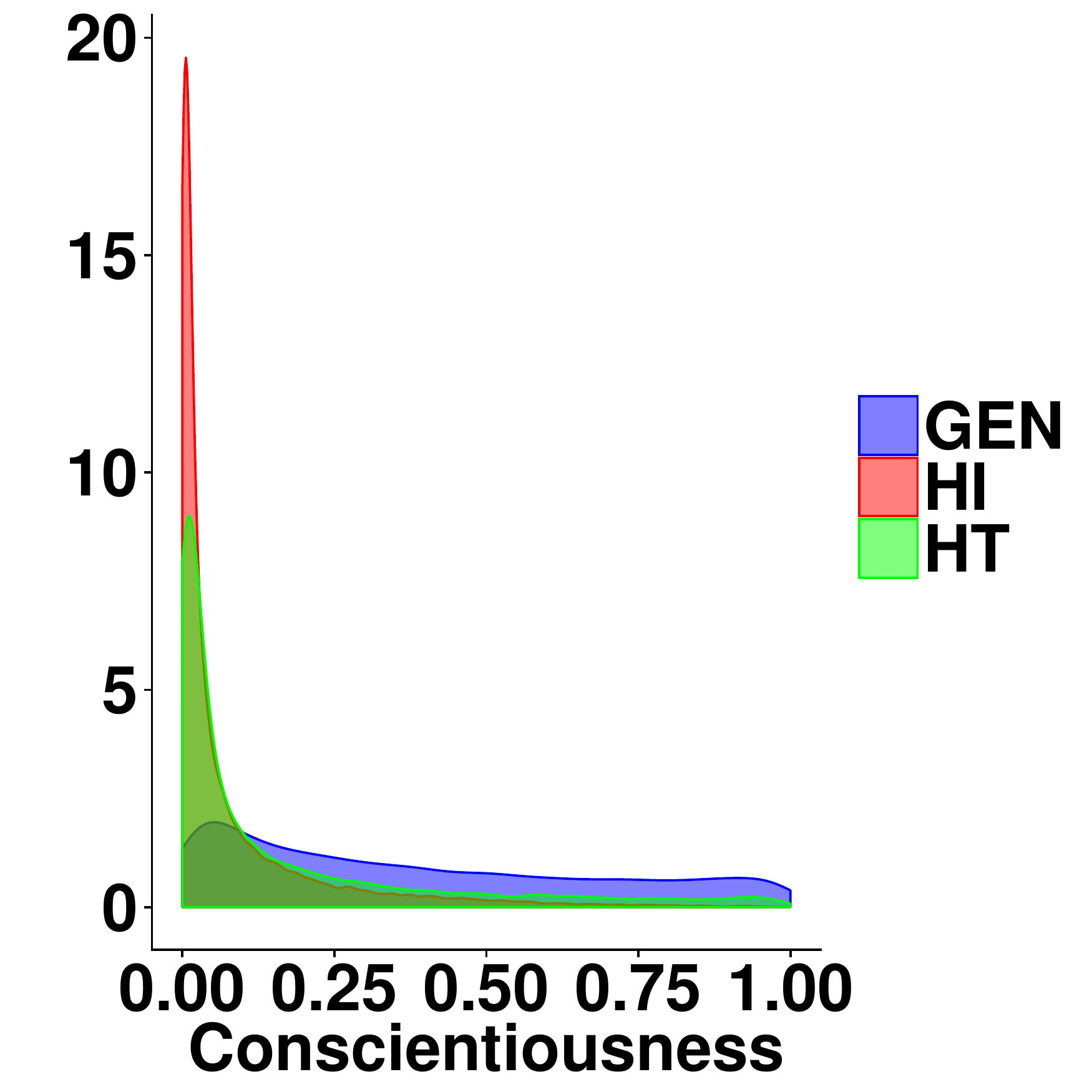}}
         {\includegraphics[width=0.19\textwidth, height = 3cm]{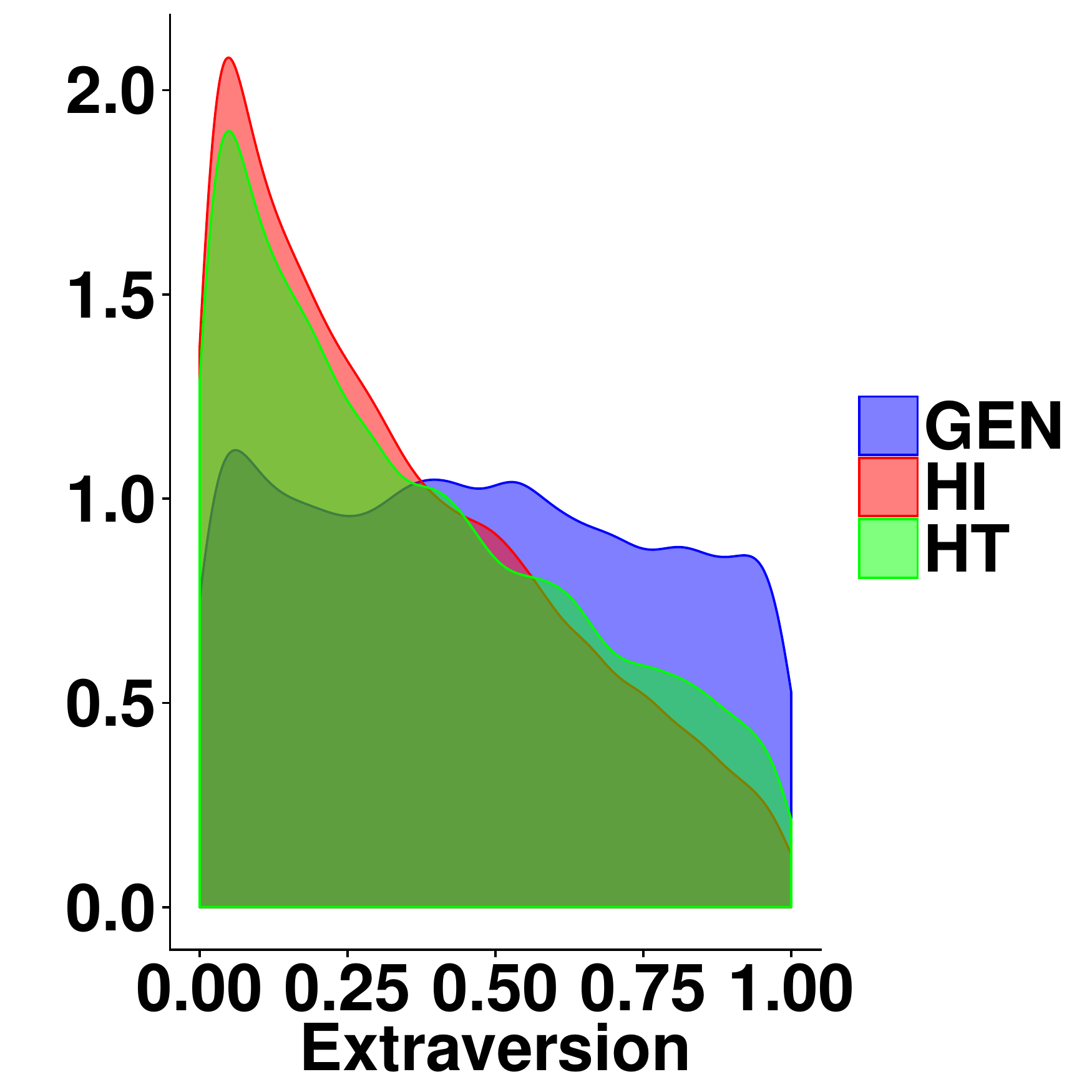}}
          \caption{Distribution of scores for the Big Five personality traits.}
  \label{fig:agree}
\end{figure*}

The IBM Watson Personality API infers personality characteristics from textual information based on an open-vocabulary approach~\cite{ibm2015science}. The API's machine learning algorithm is trained using scores obtained from surveys conducted among thousands of users along with data from their Twitter feeds.
The API provides scores [0, 1] that reflect the normalized percentile score for the characteristic. 
We analyze the results of the \emph{Big Five} personality model, the most widely used model for generally describing how a person engages with the world. The model includes five primary dimensions: Agreeableness, Conscientiousness, Extraversion, Emotional range, and Openness. 
The Big Five personality traits, their associated facets, and how to interpret them are defined in detail in~\cite{ibm2015big5output}. 


To quantify the difference between the continuous distributions of different personality aspects, we compute the Hellinger distance~\cite{tanton2005encyclopedia}. The Hellinger distance between two measures $P$ and $Q$ represented by two distributions $f(x)$ and $g(x)$, respectively, is defined as: \\
 \begin{equation}
H(P, Q) =  \sqrt{\frac{1}{2} \int (\sqrt{f(x)} - \sqrt{g(x)} \:)^2 \:dx} \:,
\end{equation} 
where $H(P,Q) \in [0, 1]$. The minimum distance of 0 is achieved when $P$ and $Q$ exhibit exactly the same distributions; the maximum distance of 1 is achieved when $P$ assigns probability zero to every set to which $Q$ assigns a positive probability, and vice versa. Table~\ref{table:personality-facets} depicts the pairwise distribution distances between HIs and HTs (HI-HT), and the distance between the HI and HT distributions and the general users, (HI-Gen-1\%) and (HT-Gen-1\%), respectively. We also report the results of the Mann-Whitney U tests. 





\noindent
\textbf{HIs and HTs personalities differ from general users:}
For all the personality traits depicted in Table~\ref{table:personality-facets}, the Hellinger distance of (HI-HT) is always less than or equal to (HI-Gen-1\%) and (HT-Gen-1\%). This indicates that HIs and HTs have more similar personalities to each other than general users. This is also shown for each personality trait's median. With the exception of Openness, the median for HIs personality facets is closer to the median of HTs than Gen-1\%.  


Both HIs and HTs exhibit lower Agreeableness than  general users. 
Lower Agreeableness scores are often associated with suspicious and antagonistic behaviors~\cite{toegel2012become}. Our results indicate that HIs and HTs are more self-focused, contrary, proud, cautious of others, and can compromise morality.

While Figure~\ref{fig:agree} shows that the distributions for HIs, HTs, and general users are close (with a median of approximately 0.5), when we investigate Openness, we find discrepancies in the lower level facets: Adventurousness, Emotionality, and Imagination.
Both HIs and HTs exhibit lower scores for Emotionality and Adventurousness, and higher Imagination scores, in comparison to the general users. Moreover, HIs and HTs have similar distributions for Artistic Interests ($p=0.24$) and Liberalism ($p=0.98$). These results indicate that HIs and HTs are less emotionally aware and less adventurous with a wild imagination (lower preference to facts), and more authority challenging behavior, in comparison to the general users. 


For Emotional range, HIs and HTs have lower scores than general users across all facets. 
HIs have slightly lower scores, but still statistically significant, than HTs. The  high Emotional range scores indicate that HIs and HTs are more fiery, prone-to worry, melancholy, hedonistic, and susceptible to stress. 
Cheng \textit{et al.}~observe that negative mood increased a user's probability to engage in trolling, and that anger begets more anger~\cite{cheng2017anyone}. It seems that Emotional range facets such as Anxiety, Depression, Immoderation, and Self-consciousness are embodied more in the tweets of HIs and HTs but further work is needed to directly correlate these parameters with hate speech and online trolling.


For Conscientiousness, HIs and HTs generally have lower scores than general users. 
Consistently, HTs score slightly higher, but still statistically significant, than HIs. Our results suggest that HIs and HTs 
tend to disregard rules and obligations, as indicated by  low dutifulness scores, and would rather take action immediately than spend time deliberating a decision, as indicated by  low Cautiousness scores. As for Extraversion, HIs and HTs tend to have lower scores of Activity-level, Friendliness, and Cheerfulness but higher scores for Excitement seeking, in comparison to general users. Our results indicate that HIs and HTs 
are inclined to be less sociable, less assertive, and more solemn. 




\noindent
\textbf{HIs and HTs tend to share personality facets:} 
It is possible that the personality facets for HIs and HTs could contribute to the problem of hate speech. Our results show that indeed the personalities of HIs and HTs are much closer to each other than to the general users. Moreover, our results agree with prior work conducted for victims of bullying. Prior studies, in workplaces and schools, have shown that bullying victims tend to show depression and helplessness as a result of bullying~\cite{price1994social}. Moreover victims are described as lacking social skills, tending to show emotions, e.g., crying easily~\cite{schwartz1993emergence}, and are likely to experience anxiety, loneliness, and hyperactivity~\cite{camodeca2003links,johnson2002vulnerability}. Our work also agrees with studies that show that bullies and victims share a wide range of bully-typifying personality traits such as machiavellianism, narcissism, psychoticism, and aggression, and that bullies and victims could exchange roles~\cite{linton2013personality}. 
Interestingly, in this work we have shown that these personality signals have been mirrored from the physical world and now have a presence in the digital world as well. 


\vspace{-0.2cm}
\section{Discussion and Conclusion}

\textbf{Hate mitigation and counter speech}. 
Successful counter speech is a direct response to hateful comments aimed at influencing discourse and behavior~\cite{benesch2014countering,benesch2016successful}. 
Recently, Munger  showed that counter speech using automated bots can reduce instances of racist speech if instigators are sanctioned by a high-follower white male~\cite{munger2017tweetment}. 
If AI-powered counter speech bots are widely deployed~\cite{forbes2017bots}, a research challenge would then be how we can 
design these bots to achieve maximum impact. Prior work has shown that people respond more positively to messages tailored to their personality~\cite{hirsh2012personalized}. For instance, Myszkowski and Storme correlated Openness with product design and found that individuals with low openness scores respond to product appearance and, conversely, high openness individuals tend to focus on product aspects~\cite{myszkowski2012personality}. Our personality analyses could be used to design next generation counter speech bots of increased effectiveness. Moreover, our personality results show that 50\% of HIs and HTs score above 0.53 for the Openness to change personality facet, which may imply that counter speech could be successfully used to decrease hate speech.


\vspace*{0.05in}
\noindent
\textbf{Profile-based data collection}. 
Most common methods of data collection use hate terms and trained classifiers to classify new content as hateful or benign. Another method employs bootstrapping, which is used in~\cite{xiang2012detecting} to obtain training data by classifying Twitter accounts as either ``good'' or ``bad'' based on usage of offensive terms. All tweets from ``bad accounts'' are marked as hate speech instances. Our results could be incorporated through the use of personality scores as features to classify users. Alternatively, a user could be represented as a vector of personality facets and then compared to values for hate speech accounts. This could be especially useful for content curation for cases when the instigator is likely to engage in hate speech more than once~\cite{xiang2012detecting,chatzakou2017mean} or as features for early instigator identification~\cite{cheng2015antisocial} and implicit hate speech detection.


\vspace*{0.05in}
\noindent
\textbf{Critique of methodology and limitations}. 
There are limitations to our methodology and findings. Recent studies~\cite{tufekci2014big,morstatter2013sample} discuss common issues associated with social media analysis and the sample quality of the Twitter Streaming API. Our analysis focused on explicit hate speech and relied on keyword-based methods, which have been shown to miss instances of hateful speech~\cite{saleem2016web}. However, while  we cannot claim to have captured a complete representation of hate speech on Twitter, as our  starting point for tweet filtering was based on a set of hate terms from Hatebase, our primary objective was to investigate hate speech instigator and target accounts with a high precision dataset. We believe that our careful curation methodology achieved this end goal.


\noindent
\textbf{Conclusion.} We have presented the first comparative study of hate speech instigators, targets, and general Twitter users. We have outlined a semi-automated classification approach for curation of directed explicit hate speech. 
Our analysis yields a number of interesting and unexpected findings about actors of hate speech. For example, we found that hate instigators target more visible users and that participating in hate commentary is associated with higher visibility. 
We also showed that hate instigators and targets have unique personality characteristics that may contribute to hate speech such as anger, depression, and immoderation. We hope that our results can be used as meta-information to improve hate speech classification, detection and mitigation to combat this increasingly pervasive problem.

\bibliographystyle{aaai}
{\footnotesize
\bibliography{refs}}

\end{document}